\documentclass[prd,amsmath,amssymb,superscriptaddress,preprintnumbers,twocolumn,nofootinbib,10pt]{revtex4-1}
\pdfoutput=1
\usepackage{graphicx}
\usepackage{dcolumn}
\usepackage{bm}
\usepackage{amssymb}
\usepackage{latexsym}
\usepackage{booktabs}
\usepackage{amsmath}
\usepackage{multirow}
\usepackage{url}
\usepackage{footnote}
\usepackage{float}
\usepackage{threeparttable}
\usepackage[colorlinks=true, linkcolor=blue, citecolor=blue]{hyperref}
\usepackage[bottom]{footmisc}

\usepackage[normalem]{ulem}
\usepackage{color}
\usepackage{array}
\usepackage{enumerate}
\usepackage{adjustbox}

\usepackage{makecell}
\usepackage{diagbox}
\usepackage{epstopdf}
\usepackage{epsfig}
\usepackage{longtable}
\usepackage{supertabular}
\usepackage{algorithm}
\usepackage{pifont}
\usepackage{algorithmic}
\usepackage{changepage}
\usepackage{setspace}

\begin{document}
\title{Cosmological search for sterile neutrinos after DESI 2024}
\author{Guo-Hong Du}
\affiliation{Liaoning Key Laboratory of Cosmology and Astrophysics, College of Sciences, Northeastern University, Shenyang 110819, China}
\author{Tian-Nuo Li}
\affiliation{Liaoning Key Laboratory of Cosmology and Astrophysics, College of Sciences, Northeastern University, Shenyang 110819, China}
\author{Peng-Ju Wu}
\affiliation{School of Physics, Ningxia University, Yinchuan 750021, China}
\author{Lu Feng}
\affiliation{College of Physical Science and Technology, Shenyang Normal University, Shenyang 110034, China}
\affiliation{Liaoning Key Laboratory of Cosmology and Astrophysics, College of Sciences, Northeastern University, Shenyang 110819, China}
\author{Sheng-Han Zhou}
\affiliation{Liaoning Key Laboratory of Cosmology and Astrophysics, College of Sciences, Northeastern University, Shenyang 110819, China}
\author{Jing-Fei Zhang}
\affiliation{Liaoning Key Laboratory of Cosmology and Astrophysics, College of Sciences, Northeastern University, Shenyang 110819, China}
\author{Xin Zhang}\thanks{Corresponding author}\email{zhangxin@mail.neu.edu.cn}
\affiliation{Liaoning Key Laboratory of Cosmology and Astrophysics, College of Sciences, Northeastern University, Shenyang 110819, China}
\affiliation{MOE Key Laboratory of Data Analytics and Optimization for Smart Industry, Northeastern University, Shenyang 110819, China}
\affiliation{National Frontiers Science Center for Industrial Intelligence and Systems Optimization, Northeastern University, Shenyang 110819, China}

\begin{abstract}
The question of whether the massive sterile neutrinos exist remains a crucial unresolved issue in both particle physics and cosmology. We explore the cosmological constraints on the massive sterile neutrinos using the latest observational data, including the baryon acoustic oscillations data from DESI, the cosmic microwave background data from Planck satellite and ACT, and the 5-year Type Ia supernova data and the 3-year weak-lensing data from DES. We search for the massive sterile neutrinos within the $\Lambda$CDM, $w$CDM, and $w_0w_a$CDM models. Our analysis shows that when considering massive sterile neutrinos within the $w_0w_a\rm CDM$ model, the combined datasets allow us to infer a non-zero sterile neutrino mass at approximately $2\sigma$ confidence level. Specifically, in the $w_0w_a$CDM+Sterile model, the effective mass of sterile neutrinos and the effective number of relativistic species are constrained to be $m_{\nu,\ \mathrm{sterile}}^{\mathrm{eff}} = 0.50^{+0.33}_{-0.27} \, \mathrm{eV}$ and $N_\mathrm{eff} = 3.076^{+0.011}_{-0.017}$, respectively. However, the $\Lambda$CDM+Sterile and $w$CDM+Sterile models could not provide evidence supporting the existence of massive sterile neutrinos.
\end{abstract}

\maketitle
\section{Introduction}\label{sec1}

The neutrino oscillation phenomena observed in neutrino experiments are well explained by oscillations among the three active neutrino flavors with nonvanishing masses and mixings \cite{PhysRevLett.81.1562-Oscillation, Super-Kamiokande:2000ywb-Oscillation, PhysRevLett.89.011301-Oscillation, SNO:2001kpb-Oscillation, Olive_2014-Oscillation, dayabaycollaboration2022precision}. However, some anomalies and discrepancies observed in neutrino experiments suggest that the existence of light sterile neutrinos as an additional flavor of neutrino remains an open question in neutrino physics. Short-baseline experiments, such as LSND \cite{LSND:2001aii-SBL} and MiniBooNE \cite{MiniBooNE:2018esg-SBL}, have reported anomalies that cannot be explained by three active neutrinos, pointing to the existence of sterile neutrinos with mass-squared differences around the $\rm eV^2$ scale. Although the results from neutrino experiments such as Daya Bay \cite{DayaBay:2016lkk-noste} and IceCube \cite{IceCube:2016rnb-noste} are consistent with the absence of sterile neutrinos, the IceCube data still yield a best-fit value of $\Delta m_{41}^2 \simeq 4.5\ {\rm eV^2}$ \cite{IceCube:2020phf-m41}. Furthermore, mild discrepancies in measurements of $\Delta m_{21}^2$ between Super-Kamiokande \cite{Super-Kamiokande:2016yck-m21} and KamLAND \cite{KamLAND:2013rgu-m21} persist. Consequently, sterile neutrinos are still regarded as a viable phenomenological motivation (see, e.g., Refs.~\cite{Gariazzo:2013gua-phenomeno, Gariazzo:2014dla-phenomeno, Hagstotz:2020ukm-phenomeno, Archidiacono:2020yey-phenomeno}), even though being challenged by other neutrino experiments.

Cosmology can provide independent constraints for eV-scale sterile neutrinos. These constraints primarily come from the cosmic microwave background (CMB), big bang nucleosynthesis (BBN), and large-scale structure observations, which are influenced by sterile neutrino free-streaming effects and the increase in the effective number of neutrino relativistic species, $N_\mathrm{eff}$. The Planck collaboration reported $N_\mathrm{eff} < 3.29$ and the effective mass of sterile neutrinos $m_{\nu,\ \mathrm{sterile}}^{\mathrm{eff}} < 0.65 \, \mathrm{eV}$ in the standard $\Lambda$ cold dark matter ($\Lambda$CDM) model, utilizing the Planck CMB data combined with the baryon acoustic oscillation (BAO) measurements from the Sloan Digital Sky Survey Data Release 12 (DR12) \cite{refId0-p18cosmo}. However, introducing sterile neutrinos alone seems not favored by cosmological observations and leads to poorer data fits. The Planck 2018 results show a strong contradiction between the presence of light thermalized sterile neutrinos and CMB data \cite{refId0-p18cosmo}, and the resulting large $N_\mathrm{eff}$ is incompatible with BBN observations \cite{DiBari:2001ua-bbn-neff,Barger:2003zg-bbn-neff,Dolgov:2003sg-bbn-neff}. Therefore, additional components, such as dynamical dark energy, could be introduced to dominate over or counteract the effects of sterile neutrinos. Considering dynamical dark energy or modified gravity theory can also relax the constraints on free-streaming, potentially allowing for the existence of massive sterile neutrinos \cite{Kristiansen:2011mp-deeos, Motohashi:2010sj-fr}. The impact of dark energy on constraints for sterile neutrinos has also been widely discussed in previous studies \cite{Hamann:2011ge-de, Li:2015poa,Kumar:2016zpg, Feng:2017usu,Feng:2017nss-de,Kumar:2017dnp, Zhao:2017urm,Feng:2017mfs-de, Yang:2017amu, Feng:2021ipq-de, DiValentino:2021rjj-de, Pan:2023frx-de, Feng:2024mfx}.

Recently, the Dark Energy Spectroscopic Instrument (DESI) collaboration released its BAO data obtained in galaxy, quasar, and Lyman-$\alpha$ (Ly$\alpha$) forest observations, and the Dark Energy Survey (DES) collaboration published a new Type Ia supernova (SN) compilation, DESY5, as a component of the 5-year data release. The combination of the BAO data from DESI, SN data from DESY5, and CMB measurements from Planck satellite and the Atacama Cosmology Telescope (ACT) suggests the possibility of dynamical evolution in dark energy, with the statistical confidence reaching to $3.9\sigma$ \cite{DESI:2024uvr,DESI:2024lzq,DESI:2024mwx}. This latest result on dark energy research has garnered widespread attention and discussions (see, e.g., Refs.~\cite{Calderon:2024uwn,Wang:2024pui,Escamilla-Rivera:2024sae,DiValentino:2024xsv,Yang:2024kdo,Wang:2024dka,Gomez-Valent:2024tdb,Allali:2024anb,Qu:2024lpx,Colgain:2024xqj,Cortes:2024lgw,Wang:2024hks,Ge:2024kac,Pedrotti:2024kpn,Jiang:2024viw,Escudero:2024uea,Giare:2024gpk,Du:2024pai,Toda:2024ncp,Wu:2025vfs,Du:2025xes,Li:2025htp,Pang:2024qyh,Li:2025dwz,Ling:2025lmw,Reboucas:2024smm,Escamilla:2024ahl,Li:2024qso,Specogna:2024euz,Feng:2025mlo,Wang:2025ljj,Li:2025muv,Zhou:2025nkb,Park:2024jns,Alestas:2024eic,Wang:2024tjd,Wu:2024faw,Li:2024qus,Ye:2024ywg,Tyagi:2024cqp,Li:2025owk,Du:2025csv,Feng:2025mlo,Yao:2025twv,Zhang:2025dwu}). The high-confidence result from DESI enables us to further investigate the cosmological bounds on sterile neutrinos within the framework of a time-evolving dark-energy equation of state (EoS). Therefore, it is necessary to investigate the impact of introducing the dynamically evolving dark energy on constraints on sterile neutrinos after the latest DESI BAO data release.

Moreover, previous studies have shown that incorporating weak-lensing data from the DES 1-year (DESY1) observation can enhance the possibility of sterile neutrino existence with non-zero mass, reaching a confidence level of $2\sigma$ \cite{Pan:2023frx-de}. This is due to the weak-lensing data predict a lower level of matter clustering, consistent with the suppression of small-scale structures caused by the free-streaming effects of sterile neutrinos. Therefore, the amplitude of matter density fluctuations at the $8\ h^{-1}$ Mpc scale, characterized by $\sigma_8$, is inversely correlated with the mass and effective number of sterile neutrinos. A lower value of $\sigma_8$ allows for the presence of more massive sterile neutrinos. DES collaboration has released its 3-year (DESY3) weak-lensing data, which consists of three two-point correlation functions: cosmic shear, galaxy clustering, and the cross-correlation between source galaxy shear and lens galaxy positions \cite{DES:2021wwk-Y3,DES:2022ccp-Y3}. The DESY3 analysis is based on observations of approximately 100 million source galaxies across a 4143 $\rm deg^2$ footprint, which is around three times larger compared to DESY1. Consequently, incorporating DESY3 data into the analysis is expected to impose further influence on searching sterile neutrinos.

In this paper, we aim to explore cosmological bounds on the massive sterile neutrinos with the impact of the dynamically evolving dark energy utilizing the latest DESI BAO and DESY5 SN data, as well as the CMB data, and the DESY1 and DESY3 weak-lensing measurements. Furthermore, we compare different models by calculating the Bayes factor. Our results demonstrate the potential of latest cosmological observations in the search for massive sterile neutrinos. 

The organization of this paper is as follows. In Sec.~\ref{sec2}, we describe the methods and data utilized in this work. In Sec.~\ref{sec3}, we report the results and provide a detailed discussion. In Sec.~\ref{sec4}, we present the conclusions of this work.

\section{Methodology and data}\label{sec2}

\begin{table*}[htbp]
\renewcommand\arraystretch{1.7}
\centering
\caption{Cosmological models and parameters used in this work.}
\setlength{\tabcolsep}{14pt}
\resizebox{0.9\textwidth}{!}{%
\begin{tabular}{lcl}
\hline\hline
Model & Parameter \# & Parameters \\ 
\hline
$\Lambda$CDM+Sterile & 8 & $\ln(10^{10} A_\mathrm{s})$, $n_\mathrm{s}$, $100\theta_\mathrm{MC}$, $\Omega_\mathrm{b} h^2$, $\Omega_\mathrm{c} h^2$, $\tau_\mathrm{reio}$, $N_\mathrm{eff}$, $m_{\nu,\ \mathrm{sterile}}^\mathrm{eff}$ \\ 
$w$CDM+Sterile & 9 & $\ln(10^{10} A_\mathrm{s})$, $n_\mathrm{s}$, $100\theta_\mathrm{MC}$, $\Omega_\mathrm{b} h^2$, $\Omega_\mathrm{c} h^2$, $\tau_\mathrm{reio}$, $N_\mathrm{eff}$, $m_{\nu,\ \mathrm{sterile}}^\mathrm{eff}$, $w$ \\ 
$w_0w_a$CDM+Sterile & 10 & $\ln(10^{10} A_\mathrm{s})$, $n_\mathrm{s}$, $100\theta_\mathrm{MC}$, $\Omega_\mathrm{b} h^2$, $\Omega_\mathrm{c} h^2$, $\tau_\mathrm{reio}$, $N_\mathrm{eff}$, $m_{\nu,\ \mathrm{sterile}}^\mathrm{eff}$, $w_0$, $w_a$ \\ 
\hline
\end{tabular}
}
\label{tab:parms}
\end{table*}

In this section, we first describe the cosmological models analyzed in this study, followed by a brief overview of the cosmological datasets utilized. Finally, we present the methods used to infer cosmological parameters and calculate Bayes factors.

\subsection{Models}\label{sec2.1}

In this analysis, we consider three cosmological models, with their respective parameters detailed in Table~\ref{tab:parms}.

\begin{itemize}

\item $\Lambda$CDM+Sterile. The baseline model is the standard $\Lambda$CDM model. This model contains six base parameters, including the logarithm of the amplitude of the primordial scalar power spectrum $\ln(10^{10} A_\mathrm{s})$, the spectral index of the primordial scalar perturbations $n_\mathrm{s}$, the angular scale of the sound horizon at decoupling $100\theta_\mathrm{MC}$, the baryon density parameter $\Omega_\mathrm{b} h^2$, the cold dark matter density parameter $\Omega_\mathrm{c} h^2$, and the reionization optical depth $\tau_\mathrm{reio}$. Dark energy in this model is described by the cosmological constant $\Lambda$, with the dark-energy EoS parameter $w = -1$. To incorporate the effects of sterile neutrinos, as described in Ref. \cite{Planck:2013pxb}, we extend the $\rm \Lambda CDM$ model by introducing $N_\mathrm{eff}$ and the effective mass of sterile neutrinos, $m_{\nu,\ \mathrm{sterile}}^\mathrm{eff}$, given by
\begin{equation}
m_{\nu,\ \mathrm{sterile}}^\mathrm{eff} \equiv (\Delta N_\mathrm{eff})^{3/4}~m_{\nu,\ \mathrm{sterile}}^\mathrm{thermal}.
\end{equation}
Here, $\Delta N_\mathrm{eff} \equiv N_\mathrm{eff} - N_\mathrm{SM}$, where $N_\mathrm{SM} = 3.044$ is the effective number of active neutrinos in the Standard Model (SM) framework \cite{Akita:2020szl, Froustey:2020mcq, Bennett:2020zkv}. $m_{\nu,\ \mathrm{sterile}}^\mathrm{thermal}$ represents the physical mass of thermally produced sterile neutrinos. In our analysis, we set a prior $m_{\nu,\ \mathrm{sterile}}^\mathrm{thermal} < 10 \, \mathrm{eV}$ as is restricted in Ref.~\cite{refId0-p18cosmo}. In addition, for the total mass of the three active neutrino species, we assume $\sum m_\nu=0.06$~eV.

\item $w$CDM+Sterile. Building on $\Lambda$CDM+Sterile model, we further consider a scenario where $w$ is treated as a free-to-vary constant.

\item $w_0w_a$CDM+Sterile. Extending $\Lambda$CDM+Sterile further, we consider that $w$ evolves with the scale factor $a$ following the Chevallier-Polarski-Linder (CPL) parametrization \cite{Chevallier:2000qy, Linder:2002et}, expressed as
\begin{equation}
w(a) = w_0 + w_a (1 - a).
\end{equation}
Here, $w_0$ represents the present value of $w$, and $w_a$ indicates the evolution of $w$ with $a$.
\end{itemize}

\subsection{Cosmological datasets}\label{sec2.2}

In our analysis, we use the following cosmological datasets.
\begin{itemize}
\item CMB. This analysis utilizes comprehensive measurements of the CMB anisotropies and lensing, incorporating multiple likelihood components. Specifically, the CMB dataset used in this work include: (i) high-$\ell$ power spectra of temperature and polarization anisotropies, $C_{\ell}^{TT}$, $C_{\ell}^{TE}$, and $C_{\ell}^{EE}$ ($\ell > 30$), derived from the NPIPE PR4 Planck \texttt{CamSpec} likelihood~\cite{refId0-p18cosmo,efstathiou2019-highl,rosenberg22-highl}; (ii) low-$\ell$ ($2 \leq \ell \leq 30$) temperature anisotropy power spectrum, $C_{\ell}^{TT}$, reconstructed using the Planck \texttt{Commander} likelihood~\cite{Pagano:2019tci-lowl,Planck:2019nip}; (iii) low-$\ell$ ($2 \leq \ell \leq 30$) large-scale E-mode polarization power spectrum, $C_{\ell}^{EE}$, derived from the Planck \texttt{SimAll} likelihood~\cite{refId0-p18cosmo,Planck:2019nip}; (iv) the lensing potential power spectrum, derived from the NPIPE PR4 Planck lensing reconstruction data\footnote{The lensing likelihood is publicly available at \url{https://github.com/carronj/planck_PR4_lensing}.} \cite{Carron:2022eyg-PR4len,Carron:2022eum-PR4len} and complemented by the ACT Data Release 6 lensing likelihood\footnote{Accessible at \url{https://github.com/ACTCollaboration/act_dr6_lenslike}.}~\cite{Madhavacheril_2024-actlen, Qu_2024-actlen}.

\item DESI. The tracers in the DESI Data Release 1\footnote{Details at \url{https://data.desi.lbl.gov/doc/releases/}.} include the bright galaxy sample (BGS) ($0.1 \leq z \leq 0.5$), luminous red galaxy (LRG) ($0.4 \leq z \leq 1.0$), emission line galaxy (ELG) ($0.8 \leq z \leq 1.6$), quasar (QSO) ($0.8 \leq z \leq 2.1$), and Ly$\alpha$ forest ($2.1 \leq z \leq 4.2$). We utilize 12 BAO measurements obtained from above tracers~\cite{DESI:2024uvr,DESI:2024lzq,DESI:2024mwx}. The dataset characterizes the large-scale structure of the universe through key distance measurements: the transverse comoving distance, $D_{\mathrm{M}}/r_{\mathrm{d}}$, the angle-averaged distance, $D_{\mathrm{V}}/r_{\mathrm{d}}$, and the Hubble horizon, $D_{\mathrm{H}}/r_{\mathrm{d}}$, where $r_{\mathrm{d}}$ is the comoving sound horizon at the drag epoch.

\item DESY5. We incorporate 1829 supernovae (SNe) data\footnote{Data provided in \url{https://github.com/des-science/DES-SN5YR}.} from the 5-year data release of DES, covering the redshift range $0.025 < z < 1.3$ \cite{descollaboration2024dark-DESY5}. This dataset comprises 1635 photometrically classified SN Ia homogeneously selected within $0.1 < z < 1.3$, complemented by 194 low-redshift SNe ($0.025 < z < 0.1$) that overlap with the Pantheon+ sample \cite{Brout:2022vxf-SN+}.

\item DESY3. We incorporate the DES 3-year release data, which combines angular galaxy clustering and weak gravitational lensing into the ``3×2pt'' datavector, consisting of three two-point correlation functions: galaxy-galaxy, galaxy-shear, and shear-shear \cite{DES:2021wwk-Y3,DES:2022ccp-Y3}. For the fiducial $\mathtt{MagLim}$ sample, the DESY3 analysis is based on observations of about 100 million source galaxies and 10 million lens galaxies over a 4143 $\mathrm{deg}^2$ footprint. Source galaxies are divided into four tomographic bins ($0 < z < 2.0$), while lens galaxies are split into six redshift bins ($0.20 < z < 1.05$), with the two highest redshift lens samples excluded from the fiducial analysis. Galaxy angular clustering is derived from lens galaxy positions, cosmic shear from the shear of source galaxies, and galaxy-shear correlations from the cross-correlation of source galaxy shear and lens galaxy positions. The dataset also includes the shear-ratio at small scales. The DESY3 analysis mitigates scale-dependent bias, baryon feedback, and nonlinearities through scale cuts and halofit-based theoretical modeling, marginalizing over astrophysical nuisance parameters \cite{Smith:2002dz-halofit,Takahashi:2012em-halofit}.

\item DESY1. For comparative analysis, we also employ the DES 1-year data, which is based on the analysis for 26 million source galaxies and 6.5 million lens galaxies over a 1321 $\mathrm{deg}^2$ footprint \cite{DES:2017myr-desy1,DES:2018zzu-desy1}.
\end{itemize}
Note that we denote the combined datasets of CMB, DESI BAO, and DESY5 SN as ``\textbf{CBS}'' in our analysis.

\begin{table*}[t]
\renewcommand\arraystretch{1.8}
\centering
\caption{The 1$\sigma$ or 2$\sigma$ confidence ranges of cosmological parameters obtained by using the CBS, CBS+DESY1, and CBS+DESY3 data combinations, for the $\Lambda$CDM+Sterile, $w$CDM+Sterile, and $w_0w_a$CDM+Sterile models. For the parameters $N_{\rm eff}$ and $m_{\nu,\ \rm sterile}^{\rm eff}$, when central values cannot be determined, we provide their 2$\sigma$ upper limits. For other parameters, the 1$\sigma$ confidence intervals are reported.}
\setlength{\tabcolsep}{6pt}
\label{table: St}
\resizebox{1.0\textwidth}{!}{%
\begin{tabular}{lccccccc}
\hline\hline
Parameter  & $H_0\ [\rm km\ s^{-1}\ Mpc^{-1}]$ & $\Omega_{\rm m}$ & $S_8$ & $w\ {\rm or}\ w_0$ & $w_a$ & $m_{\nu,\ \rm sterile}^{\rm eff}\ [\rm eV]$ & $N_{\rm eff}$ \\
\hline
\multicolumn{8}{l}{$\boldsymbol{\Lambda{\rm CDM}+{\rm Sterile}}$} \\
CBS          & $68.10^{+0.50}_{-0.93}$     & $0.3110 \pm 0.0054$ & $0.821^{+0.012}_{-0.008}$ & --- & --- & $< 0.217$ & $< 3.42$  \\
CBS+DESY1    & $68.03^{+0.32}_{-0.67}$     & $0.3087 \pm 0.0052$ & $0.806^{+0.014}_{-0.010}$ & --- & --- & $< 0.436$ & $3.121^{+0.004}_{-0.074}$  \\
CBS+DESY3    & $67.87^{+0.27}_{-0.48}$     & $0.3093 \pm 0.0046$ & $0.796^{+0.021}_{-0.015}$ & --- & --- & $< 0.801$ & $3.099^{+0.009}_{-0.047}$ \\
\hline
\multicolumn{8}{l}{$\boldsymbol{w{\rm CDM}+{\rm Sterile}}$} \\
CBS          & $ 67.57^{+0.85}_{-1.00} $ & $ 0.3163 \pm 0.0066 $ & $ 0.823^{+0.011}_{-0.008} $ & $-0.965\pm 0.024$ & --- & $ < 0.160 $ & $ <3.50$ \\
CBS+DESY1    & $ 67.45^{+0.67}_{-0.97} $ & $ 0.3138 \pm 0.0065 $ & $ 0.809^{+0.013}_{-0.010} $ & $-0.965 \pm 0.024$ & --- & $ < 0.445 $ & $ 3.153^{+0.035}_{-0.108} $ \\
CBS+DESY3    & $ 67.03^{+0.62}_{-0.78} $ & $ 0.3160 \pm 0.0062 $ & $ 0.797^{+0.023}_{-0.015} $ & $ -0.962 \pm 0.023 $ & --- & $ < 0.806 $ & $ 3.106^{+0.015}_{-0.056} $ \\
\hline
\multicolumn{8}{l}{$\boldsymbol{w_0w_a{\rm CDM}+{\rm Sterile}}$} \\
CBS          & $ 67.57^{+0.71}_{-0.85} $ & $ 0.3172 \pm 0.0067 $ & $ 0.825^{+0.013}_{-0.010} $ & $ -0.718 \pm 0.070 $ & $ -1.12^{+0.36}_{-0.29} $ & $<0.35$ & $ <3.37$ \\
CBS+DESY1    & $ 67.64^{+0.68}_{-0.76} $ & $ 0.3154 \pm 0.0067 $ & $ 0.812^{+0.013}_{-0.011} $ & $ -0.703 \pm 0.069 $ & $ -1.20^{+0.35}_{-0.30} $ & $ 0.19^{+0.06}_{-0.17} $ & $ 3.142^{+0.023}_{-0.090} $ \\
CBS+DESY3    & $ 67.11 \pm 0.65 $ & $ 0.3161 \pm 0.0064 $ & $ 0.793^{+0.019}_{-0.021} $ & $ -0.735 \pm 0.066 $ & $ -0.98^{+0.31}_{-0.27} $ & $ 0.50^{+0.33}_{-0.27} $ & $ 3.076^{+0.011}_{-0.017} $ \\
\hline
\end{tabular}
}
\end{table*}

\subsection{Bayes analysis}\label{sec2.3}

We perform Bayesian inference to constrain cosmological parameters listed in Table~\ref{tab:parms}. It is worth noting that in addition to these parameters, we also focus on the derived parameters $S_8\equiv\sigma_8(\Omega_{\rm m}/0.3)^{0.5}$, as weak-lensing data are included in our analysis. Theoretical calculations for cosmological observables are performed using the Boltzmann solver $\mathtt{CAMB}$\footnote{\url{https://github.com/cmbant/CAMB}} \cite{Lewis:1999bs-camb,Howlett:2012mh-camb}, which computes model predictions based on input parameters. Bayesian inference is implemented with the cosmological analysis framework $\mathtt{Cobaya}$\footnote{\url{https://github.com/CobayaSampler/cobaya}} \cite{Torrado_2021-cobaya}, incorporating likelihoods from DESY5 and DESI BAO data, as well as publicly available CMB likelihoods integrated into or compatible with $\mathtt{Cobaya}$. For cases involving the combination of Planck and ACT lensing likelihoods, we adopt higher precision settings recommended by ACT. Additionally, for DESY3 data, we utilize the likelihood from the publicly $\mathtt{CosmoLike}$ pipeline \cite{Krause:2016jvl-cosmolike}, integrated into $\mathtt{Cobaya}$ through the $\mathtt{Cocoa}$ package\footnote{\url{https://github.com/CosmoLike/cocoa}} \cite{Miranda:2020lpk-cocoa}. The consistency of this likelihood with the commonly used pipeline $\mathtt{Cosmosis}$\footnote{\url{https://github.com/joezuntz/cosmosis}} \cite{Zuntz:2014csq-cosmosis} has been thoroughly validated \cite{Miranda:2020lpk-cocoa}. Sampling is conducted using the Markov Chain Monte Carlo (MCMC) method based on the Metropolis-Hastings algorithm \cite{Lewis:2002ah-mcmc,taking-mcmc,Lewis:2013hha-mcmc}. Convergence is determined using the Gelman-Rubin criterion, and we consider chains to have converged when $R-1 < 0.01$. Finally, we analyze the MCMC results and derive parameter constraints using $\mathtt{Getdist}$\footnote{\url{https://github.com/cmbant/getdist}} \cite{lewis2019getdist}.

Furthermore, to facilitate a quantitative comparison for various models, the Bayesian evidence $Z$ is introduced as a measure of model likelihood, given by
\begin{equation}
Z \equiv P(D|M) = \int_{\Omega} P(D|\bm{\theta},M)P(\bm{\theta}|M)P(M)\ {\rm d}\bm{\theta},
\label{eq: lnZ}
\end{equation}
where $P(D|M)$ denotes the probability of the observed data $D$ given model $M$, $P(D|\bm{\theta},M)$ is the likelihood of $D$ given parameters $\bm{\theta}$ and model $M$, $P(\bm{\theta}|M)$ is the prior distribution of $\bm{\theta}$ conditioned on $M$, and $P(M)$ represents the prior of model $M$. Using the evidence $Z$, the Bayes factor in logarithmic space, $\ln \mathcal{B}_{ij}$, is calculated directly as $\ln \mathcal{B}_{ij} = \ln Z_i - \ln Z_j$, where $Z_i$ and $Z_j$ are the Bayesian evidence values for models $i$ and $j$.

The Jeffreys' scale \cite{Bayes-Jeffrey,Trotta_2008-Jeffrey} provides a generally used criterion for evaluating the relative strength of evidence between two models. According to this scale, the evidence is deemed inconclusive if $|\ln \mathcal{B}_{ij}| < 1$. For $1 \leq |\ln \mathcal{B}_{ij}| < 2.5$, the support is considered weak, while $2.5 \leq |\ln \mathcal{B}_{ij}| < 5$ indicates moderate evidence. A strong preference corresponds to $5 \leq |\ln \mathcal{B}_{ij}| < 10$, and decisive evidence is indicated by $|\ln \mathcal{B}_{ij}| \geq 10$.

The parameter space $\Omega$ involved in this work can have up to 40 dimensions, posing a significant challenge for calculating the Bayes factor. To address this, we employ a normalization flow method based on harmonic mean estimation using posterior samples, implemented with the $\mathtt{harmonic}$\footnote{\url{https://github.com/astro-informatics/harmonic}} package \cite{machine-harmonic,rzad051-harmonic}. This approach has been validated in the DESY1 analysis, demonstrating consistency with traditional nested sampling methods as well as robust performance and generalizability \cite{polanska2024learned-harmonic}.

\section{Results and Discussions}\label{sec3}

\begin{figure*}[t]
\resizebox{\textwidth}{!}{
\begin{minipage}{\textwidth}
\centering
\includegraphics[width=\linewidth]{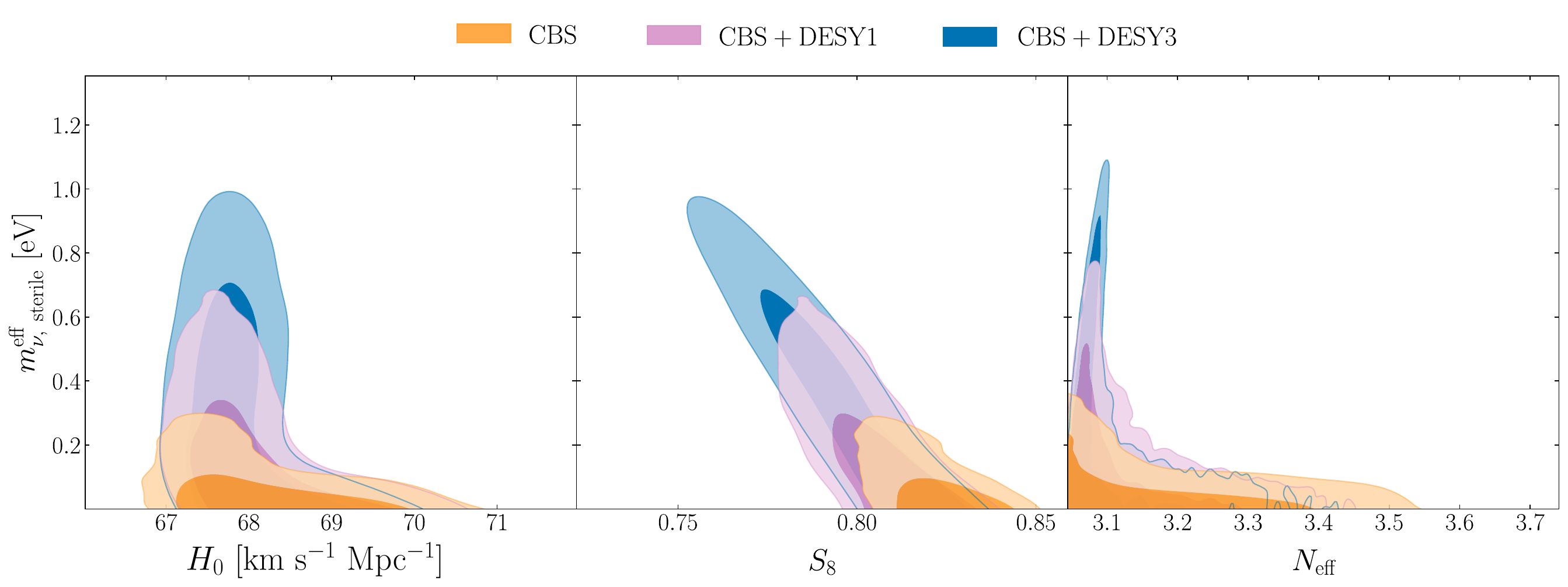}
\end{minipage}
}
\centering
\caption{The 1$\sigma$ and 2$\sigma$ credible-interval contours of $m_{\nu,\ \rm sterile}^{\rm eff}$, $H_0$, $S_8$, and $N_{\rm eff}$ in the $\Lambda$CDM+Sterile model, using the CBS, CBS+DESY1, and CBS+DESY3 data combinations, respectively.}
\label{Fig: LCDM}
\end{figure*}

\begin{figure*}[htbp]
\resizebox{\textwidth}{!}{
\begin{minipage}{\textwidth}
\centering
\includegraphics[width=\linewidth]{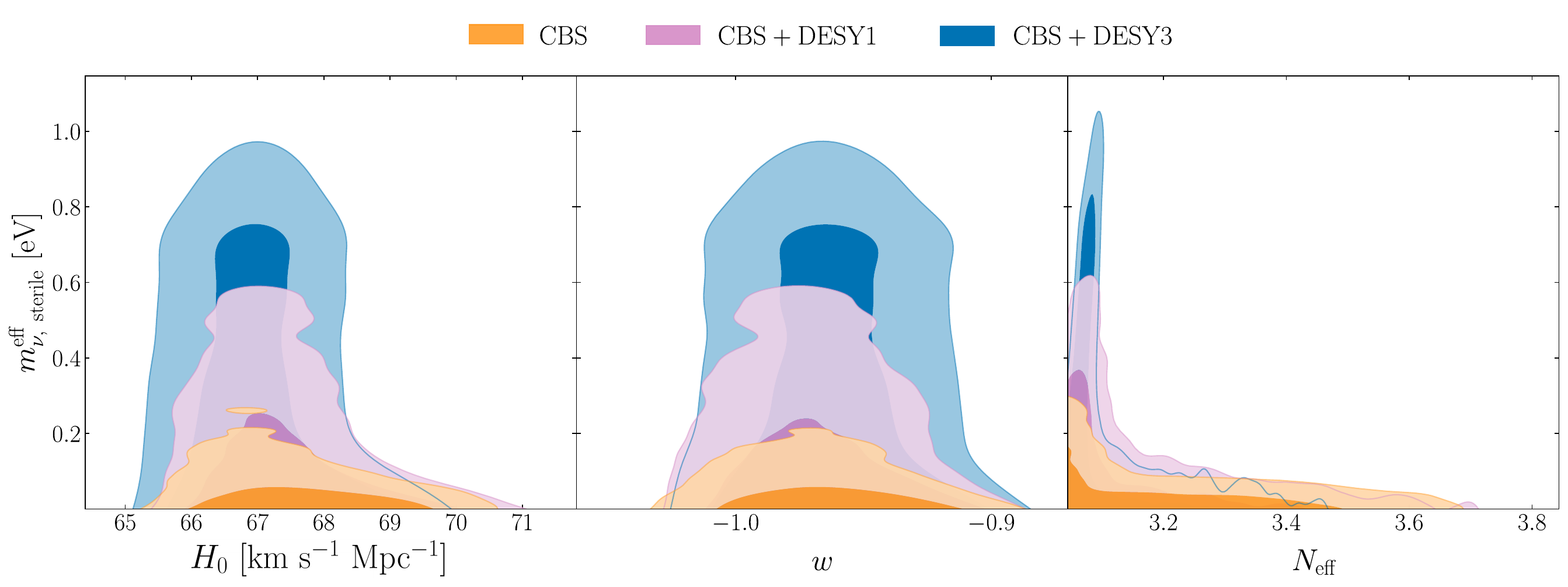}
\end{minipage}
}
\centering
\caption{The 1$\sigma$ and 2$\sigma$ credible-interval contours of $m_{\nu,\ \rm sterile}^{\rm eff}$, $H_0$, $w$, and $N_{\rm eff}$ in the $w$CDM+Sterile model, using the CBS, CBS+DESY1, and CBS+DESY3 data combinations, respectively.}
\label{Fig: WCDM}
\end{figure*}

\begin{figure*}[htbp]
\resizebox{\textwidth}{!}{
\begin{minipage}{\textwidth}
\centering
\includegraphics[width=\linewidth]{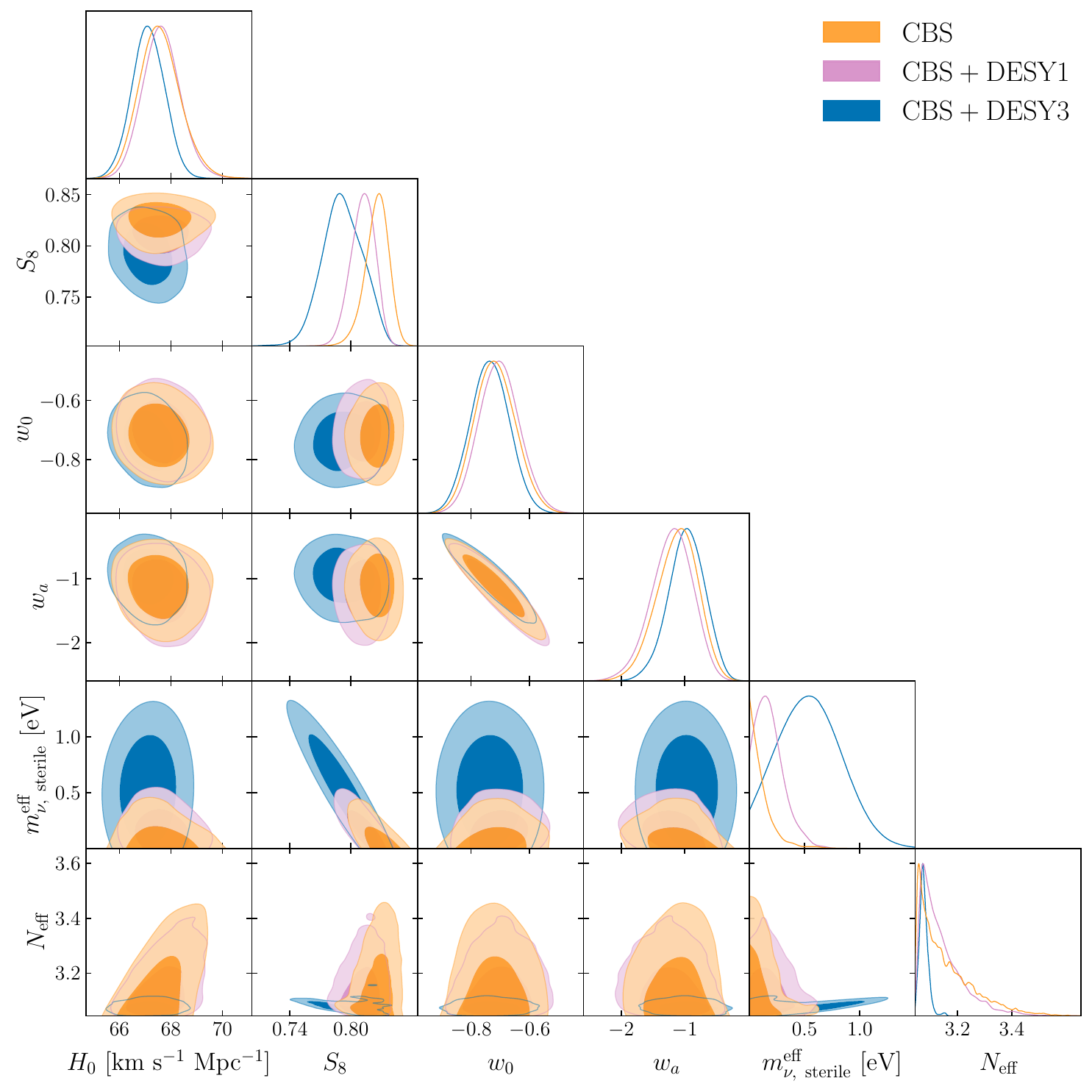}
\end{minipage}
}
\centering
\caption{The triangular plots of the marginalized posterior distributions for cosmological parameters in the $w_0w_a$CDM+Sterile model, using the CBS, CBS+DESY1, and CBS+DESY3 data combinations, respectively.}
\label{Fig: CPL}
\end{figure*}

\begin{figure}[H]
\resizebox{0.5\textwidth}{!}{
\begin{minipage}{\textwidth}
\centering
\includegraphics[width=\linewidth]{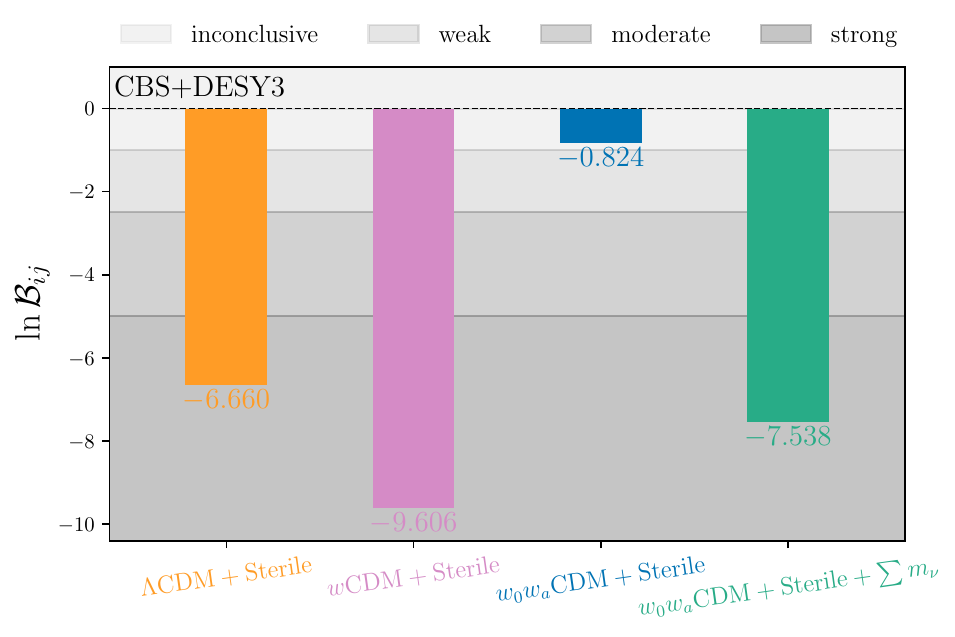}
\end{minipage}
}
\centering
\caption{{The Bayes factors of the $\Lambda$CDM+Sterile, $w$CDM+Sterile, $w_0w_a$CDM+Sterile, and $w_0w_a$CDM+Sterile+$\sum m_\nu$ models, relative to the $\Lambda$CDM model, calculated using the CBS+DESY3 data (marked in magenta, orange, blue, and green respectively).} The shades of gray from light to dark represent inconclusive, weak, moderate, and strong evidence based on Jeffreys' scale. The negative values of $\ln \mathcal{B}_{ij}$ indicate that the $\Lambda$CDM model is preferred.}
\label{Fig: BF}
\end{figure}

In this section, we present and analyze the constraints on cosmological parameters. We analyze three dark-energy models with sterile neutrinos: $\Lambda$CDM+Sterile, $w$CDM+Sterile, and $w_0w_a$CDM+Sterile. The analysis utilizes the CBS, CBS+DESY1, and CBS+DESY3 datasets. Table~\ref{table: St} summarizes the marginalized 1$\sigma$ or 2$\sigma$ (for the parameters $N_{\rm eff}$ and $m_{\nu,\ \rm sterile}^{\rm eff}$ when central values cannot be determined) constraints for the parameters of three models. Figs.~\ref{Fig: LCDM} and \ref{Fig: WCDM} show the posterior distribution contours for selected parameters of the $\Lambda$CDM+Sterile and $w$CDM+Sterile models, respectively. We also present the triangular plots of the marginalized posterior distributions of the $w_0w_a$CDM+Sterile model in Fig.~\ref{Fig: CPL}. Finally, we compare the models using the Bayes factors, as shown in Fig.~\ref{Fig: BF}.

In Fig.~\ref{Fig: LCDM}, we show the two-dimensional (2D) contour plots of $m_{\nu,\ \rm sterile}^{\rm eff}$, $H_0$, $S_8$, and $N_\mathrm{eff}$ for the $\Lambda$CDM+Sterile model. The 2$\sigma$ upper limits on the effective mass of sterile neutrinos are $m_{\nu,\ \rm sterile}^{\rm eff} < 0.217$ eV, $m_{\nu,\ \rm sterile}^{\rm eff} < 0.406$ eV, and $m_{\nu,\ \rm sterile}^{\rm eff} < 0.801$ eV using the CBS, CBS+DESY1, and CBS+DESY3 datasets, respectively. The inclusion of DESY3 significantly expands the allowed mass range for sterile neutrinos compared to CBS alone. As shown in Fig.~\ref{Fig: LCDM}, this is due to the anti-correlation of $S_8$ and $m_{\nu,\ \rm sterile}^{\rm eff}$, where CBS+DESY3 yields a lower value of $S_8 = 0.796^{+0.021}_{-0.015}$. Meanwhile, considering the free-streaming effects of sterile neutrinos could weaken the matter clustering, the $S_8$ value is lower than that in the $\Lambda$CDM model when using the same dataset. Additionally, the constraints on $N_\mathrm{eff}$ are $N_\mathrm{eff} < 3.42$, $N_\mathrm{eff} = 3.121^{+0.004}_{-0.074}$, and $N_\mathrm{eff} = 3.099^{+0.009}_{-0.047}$ using CBS, CBS+DESY1, and CBS+DESY3, respectively. This suggests that the inclusion of DESY1 and DESY3 allows for certain effective degrees of freedom for sterile neutrinos, although the results remain conservative overall. 

In Fig.~\ref{Fig: WCDM}, we present the 2D contours of $m_{\nu,\ \rm sterile}^{\rm eff}$, $H_0$, $w$, and $N_\mathrm{eff}$ for the $w$CDM+Sterile model. It can be seen that the best-fit values of $m_{\nu,\ \rm sterile}^{\rm eff}$ and $N_\mathrm{eff}$ in the $w$CDM+Sterile model are generally consistent with those in the $\Lambda$CDM+Sterile model when using the same datasets, as shown in right panels of Figs.~\ref{Fig: LCDM} and \ref{Fig: WCDM}. This indicates that the introduction of a constant dark-energy EoS has little impact on the properties of sterile neutrinos. Moreover, CBS, CBS+DESY1, and CBS+DESY3 yield generally consistent constraints on the dark-energy EoS parameter, with $w = -0.962\pm0.023$. The result $w > -1$ leads to a lower value of $H_0$ compared to the $\Lambda$CDM+Sterile model.

We present the triangular plot of the parameters in the $w_0w_a$CDM+Sterile model in Fig.~\ref{Fig: CPL}. For the effective mass of sterile neutrinos, CBS alone provides only an upper limit of $m_{\nu,\ \rm sterile}^{\rm eff} < 0.35 \, \mathrm{eV}$, while CBS+DESY1 gives $m_{\nu,\ \rm sterile}^{\rm eff} = 0.19^{+0.06}_{-0.17} \, \mathrm{eV}$, indicating a non-zero sterile neutrino mass at approximately $1\sigma$ level. In comparison with Ref.~\cite{Pan:2023frx-de}, which reports $m_{\nu,\ \rm sterile}^{\rm eff} = 0.40^{+0.16}_{-0.20} \, \mathrm{eV}$ using Planck CMB+DR12 BAO+Pantheon SN+DESY1 datasets, the latest CBS data lower the sterile neutrino mass by approximately 50\%. A notable result is that CBS+DESY3 yields $m_{\nu,\ \rm sterile}^{\rm eff} = 0.50^{+0.33}_{-0.27} \, \mathrm{eV}$, inferring a non-zero sterile neutrino mass at approximately $2\sigma$ confidence level. The best-fit value for the sterile neutrino mass is also higher than the one reported in Ref.~\cite{Pan:2023frx-de}. On one hand, this outcome can be attributed to the lower value of $S_8 = 0.793^{+0.019}_{-0.021}$ given by CBS+DESY3, which allows for a higher sterile neutrino mass. On the other hand, it reflects the consideration of a time-varying dark-energy EoS, which dominates over or counteracts the impact of a non-zero sterile neutrino mass, as discussed in Sec.~\ref{sec1}. Regarding the dark-energy EoS parameters, CBS alone gives $w_0 = -0.718 \pm 0.070$ and $w_a = -1.12^{+0.36}_{-0.29}$, while CBS+DESY1 yields $w_0 = -0.703 \pm 0.069$ and $w_a = -1.20^{+0.35}_{-0.30}$. CBS+DESY3, in contrast, provides $w_0 = -0.735 \pm 0.066$ and $w_a = -0.98^{+0.31}_{-0.27}$. Incorporating DESY1 leads to a higher $w_0$ and a smaller $w_a$ compared to CBS alone, while combining CBS with DESY3 shows the opposite trend. This indicates that a larger non-zero sterile neutrino mass tends to weaken the evolution of the dark-energy EoS.

It is emphasized that using CBS+DESY3, we obtain $N_{\mathrm{eff}} = 3.076^{+0.011}_{-0.017}$ and $\Delta N_{\mathrm{eff}} = 0.032^{+0.011}_{-0.017}$, which implies $\Delta N_{\mathrm{eff}} > 0$ at the $2\sigma$ confidence level. The value of $\Delta N_{\rm eff}$ we obtain is much less than 1, which indicates that the sterile neutrinos are only partially thermalized or has not fully reached thermal equilibrium. This incomplete thermalization can be explained by various theoretical mechanisms. For example, a small active-sterile mixing angle, a large mass splitting, or a significant lepton asymmetry can each suppress the thermalization of sterile neutrinos, thereby leading to a low value of $\Delta N_{\rm eff}$~\cite{Abazajian:2012ys, Hannestad:2012ky, Dolgov:2002wy, Alonso-Alvarez:2022uxp}. {Moreover, to account for the influence of active neutrinos on the evidence for sterile neutrinos, we perform an additional analysis within the $w_0w_a$CDM model. This extended model treats the total active neutrino mass $\sum m_\nu$, $m_{\nu,\ \rm sterile}^{\rm eff}$, and $N_{\mathrm{eff}}$ as free parameters, and is labeled as the $w_0w_a$CDM+Sterile+$\sum m_\nu$ model. The results indicate that allowing $\sum m_\nu$ to vary removes the $2\sigma$ evidence for a massive sterile neutrino and for $\Delta N_{\rm eff}>0$, leaving only upper limits on $m_{\nu,\,\rm sterile}^{\rm eff}$ and on $N_{\rm eff}$. Detailed results and discussions regarding this additional analysis are provided in Appendix~\ref{appendix1}.}

Finally, we employ Bayesian evidence to compare different models. {Figure~\ref{Fig: BF} illustrates the Bayes factors for the $\Lambda$CDM+Sterile, $w$CDM+Sterile, $w_0w_a$CDM+Sterile, and $w_0w_a$CDM+Sterile+$\sum m_\nu$ models relative to $\Lambda$CDM, based on the CBS+DESY3 dataset. The Bayes factors for the $\Lambda$CDM+Sterile, $w$CDM+Sterile, $w_0w_a$CDM+Sterile, and $w_0w_a$CDM+Sterile+$\sum m_\nu$ models are $-6.660$, $-9.606$, $-0.824$, and $-7.538$, respectively. This result indicates that introducing massive sterile neutrinos in the $\Lambda$CDM model alone is disfavored by the latest data, consistent with the findings of Planck \cite{refId0-p18cosmo}. Similarly, the $w$CDM+Sterile model is disfavored with almost decisive evidence, in line with the previous study \cite{Du:2024pai}. However, if we further consider a time-evolving dark-energy EoS, the preference of the $w_0w_a$CDM+Sterile model becomes comparable to that of the $\Lambda$CDM model based on latest data. This highlights the significant impact of dark-energy nature on the search for massive sterile neutrinos. Furthermore, when both massive sterile neutrinos and active neutrinos are considered in the $w_0w_a$CDM model, this model is again disfavored with strong evidence. This suggests that the treatment of active neutrinos as free parameters has a substantial impact on the statistical evidence for the existence of massive sterile neutrinos.}

\begin{table*}[!htbp]
\renewcommand\arraystretch{1.8}
\centering
\caption{{The 1$\sigma$ or 2$\sigma$ confidence ranges of cosmological parameters for the $w_0w_a$CDM+Sterile+$\sum m_\nu$ model obtained by the CBS, CBS+DESY1, and CBS+DESY3 data.}}
\setlength{\tabcolsep}{6pt}
\label{table3}
\resizebox{1.0\textwidth}{!}{%
\begin{tabular}{lcccccccc}
\hline\hline
Parameter  & $H_0\ [\rm km\ s^{-1}\ Mpc^{-1}]$ & $\Omega_{\rm m}$ & $S_8$ & $w_0$ & $w_a$ & $m_{\nu,\ \rm sterile}^{\rm eff}\,[\rm eV]$ & $\sum m_\nu\,[\mathrm{eV}]$ & $N_{\rm eff}$ \\
\hline
\multicolumn{9}{l}{$\boldsymbol{w_0w_a{\rm CDM}+{\rm Sterile}+\sum m_\nu}$} \\
CBS          & $ 71.34\pm 0.70 $ & $ 0.3127\pm 0.0070 $ & $ 0.8383\pm 0.0098 $ & $ -0.733^{+0.064}_{-0.074} $ & $ -0.97^{+0.39}_{-0.31} $ & $<0.25$ & $ <0.24 $ & $ <4.43$ \\
CBS+DESY1    & $ 71.28\pm 0.70 $ & $ 0.3109\pm 0.0070 $ & $ 0.8311\pm 0.0097 $ & $ -0.708^{+0.067}_{-0.078} $ & $ -1.11^{+0.41}_{-0.34} $ & $<0.26$ & $ <0.24 $ & $ <4.44$ \\
CBS+DESY3    & $ 71.49\pm 0.69 $ & $ 0.3121\pm 0.0068 $ & $ 0.8250\pm 0.0085 $ & $ -0.744^{+0.063}_{-0.073} $ & $ -0.88^{+0.37}_{-0.29} $ & $<0.37$ & $ <0.25 $ & $ <4.33$ \\
\hline
\end{tabular}
}
\end{table*}

\begin{figure*}[htbp]
\resizebox{\textwidth}{!}{
\begin{minipage}{\textwidth}
\centering
\includegraphics[width=\linewidth]{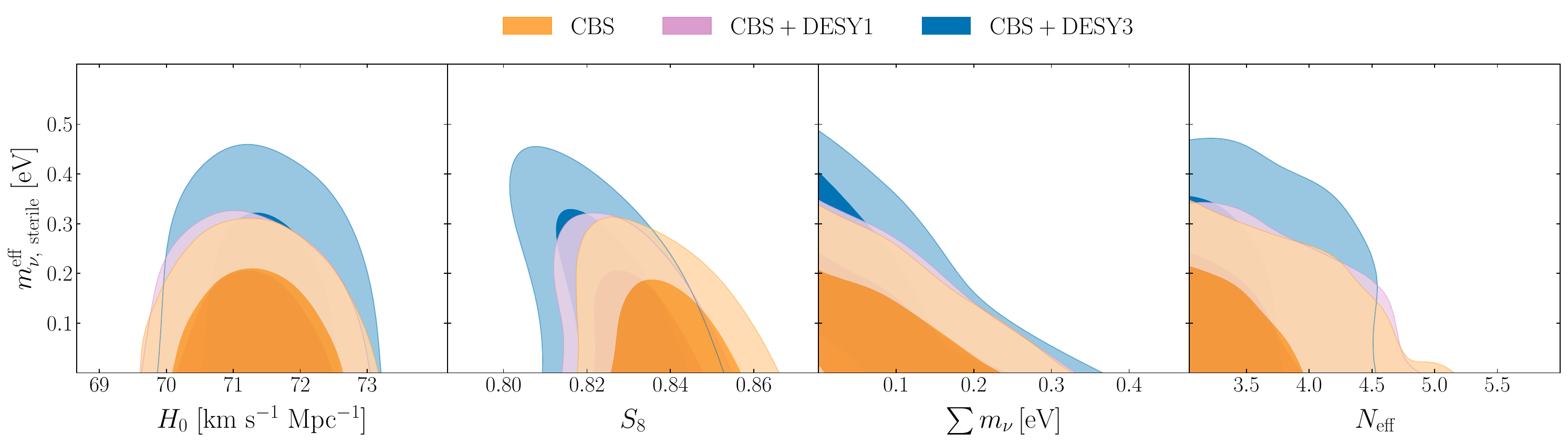}
\end{minipage}
}
\centering
\caption{{The 1$\sigma$ and 2$\sigma$ credible-interval contours of $m_{\nu,\ \rm sterile}^{\rm eff}$, $H_0$, $S_8$, $\sum m_\nu$, and $N_{\rm eff}$ in the $w_0w_a$CDM+Sterile+$\sum m_\nu$ model, using the CBS, CBS+DESY1, and CBS+DESY3 datasets, respectively.}}
\label{fig5}
\end{figure*}

\section{Conclusion}\label{sec4}

In this work, we investigate the constraints on sterile neutrinos by considering three cosmological models, including the $\Lambda$CDM+Sterile, $w$CDM+Sterile, and $w_0w_a$CDM+Sterile models. Our primary aim is to search for massive sterile neutrinos with particular focus on the $w_0w_a$CDM+Sterile model, using the latest DESI BAO, DESY5 SN, and DESY1 and DESY3 weak-lensing data.

The results from $\Lambda$CDM+Sterile and $w$CDM+Sterile models consistently reveal that incorporating the DES weak-lensing datasets increases the upper limits of sterile neutrino mass, especially combining with DESY3 due to the anti-correlation between $S_8$ and $m_{\nu,\mathrm{sterile}}^{\mathrm{eff}}$. For instance, the $\Lambda$CDM+Sterile model provides an upper limit on the sterile neutrino mass of $m_{\nu,\mathrm{sterile}}^{\mathrm{eff}} < 0.801 \, \mathrm{eV}$ using CBS+DESY3 dataset. Moreover, the introduction of a constant dark-energy EoS parameter $w$ has essentially no impact on the constraints on sterile neutrinos compared to $w=-1$.

For the $w_0w_a$CDM+Sterile model, CBS+DESY3 dataset provides approximately $2\sigma$ confidence for the existence of sterile neutrinos due to the consideration of a time-varying dark-energy EoS. Speciﬁcally, CBS+DESY3 infers the sterile neutrino effective mass of $m_{\nu,\mathrm{sterile}}^{\mathrm{eff}} = 0.50^{+0.33}_{-0.27} \, \mathrm{eV}$, achieving approximately $2\sigma$ detection confidence. Meanwhile, we also obtain $N_{\mathrm{eff}} = 3.076^{+0.011}_{-0.017}$ and $\Delta N_{\mathrm{eff}} > 0$ at the $2\sigma$ level, suggesting the thermalization of sterile neutrinos. Additionally, the constraints on the dark-energy EoS parameters, $w_0 = -0.735 \pm 0.066$ and $w_a = -0.98^{+0.31}_{-0.27}$, indicate a shift toward less dynamical dark-energy behavior when DESY3 is included. Although incorporating massive sterile neutrinos alone within the $\Lambda$CDM model is not favored by current data, the Bayesian evidence analysis further demonstrates comparable favor for the $w_0w_a$CDM+Sterile model to the $\Lambda$CDM model with the CBS+DESY3 dataset.

These findings underscore the significance of combining CMB and large-scale structure datasets to search for massive sterile neutrinos within the framework of various dark-energy models. Future surveys with improved precision are expected to provide stronger evidence for these phenomena.

\begin{acknowledgments}
We are grateful to Jun-Da Pan for helpful discussions. This work was supported by the National SKA Program of China (Grants Nos. 2022SKA0110200 and 2022SKA0110203), the National Natural Science Foundation of China (Grants Nos. 12533001, 12575049, 12473001, and 12305069), the National 111 Project (Grant No. B16009), the China Manned Space Program (Grant No. CMS-CSST-2025-A02), and the Program of the Education Department of Liaoning Province (Grant No. JYTMS20231695).
\end{acknowledgments}

\appendix
\section{{Impact of $\sum m_\nu$ on the evidence for massive sterile neutrinos}}\label{appendix1}
{In this appendix, we present the results of including $\sum m_\nu$, $m_{\nu,\ \rm sterile}^{\rm eff}$, and $N_{\mathrm{eff}}$ as free parameters in the $w_0w_a$CDM model and analyze their constraints. The $1\sigma$ or $2\sigma$ confidence ranges for the parameters are summarized in Table~\ref{table3}, and the two-dimensional credible-interval contours of the key parameters are displayed in Fig.~\ref{fig5}. Using the CBS+DESY3 dataset, we obtain upper limits of $\sum m_\nu < 0.25$ eV, $m_{\nu,\ \rm sterile}^{\rm eff} < 0.37$ eV, and $N_{\mathrm{eff}} < 4.33$, which implies that including $\sum m_\nu$ removes the $2\sigma$ evidence for a massive sterile neutrino and $\Delta N_{\rm eff}>0$. Furthermore, the results from CBS+DESY1 are largely consistent with those from CBS, both yielding a relatively small $m_{\nu,\ \rm sterile}^{\rm eff}$ compared to CBS+DESY3, as illustrated in Fig.~\ref{fig5}. It is noteworthy that in this model, CBS+DESY3 gives $H_0 = 71.49 \pm 0.69\,\rm km\,s^{-1}\,Mpc^{-1}$, reducing the $H_0$ tension to $1.2\sigma$. This reduction may be due to the inclusion of $\sum m_\nu$, which results in a larger $N_{\mathrm{eff}}$.}

\bibliography{sterile_desi}

\end{document}